\documentclass{elsart}

 \usepackage{graphics}
 \usepackage{graphicx}
 \usepackage{epsfig}

\usepackage{amssymb}

\begin{document}

\begin{frontmatter}



\title{Absence of a True Vortex-Glass Phase above the Bragg Glass Transition Line in
Bi$_2$Sr$_2$CaCu$_2$O$_{8+\delta}$}


\author{S. L. Li and H. H. Wen}

\address{ National Laboratory
for Superconductivity, Institute of Physics and Center for
Condensed Matter Physics, Chinese Academy of Sciences, P.O. Box
603, Beijing 100080, China}

\begin{abstract}
In magnetic measurements on Bi$_2$Sr$_2$CaCu$_2$O$_{8+\delta}$
(Bi-2212) single crystals, a general peak with a dynamical feature
on both $S-H$ and $S-T$ curves was found with S the magnetic
relaxation rate. At higher fields, the characteristic exponent
$\mu$ becomes negative, together with the positive curvature of
$logE$ vs. $ logj$ and the scaling based on the 2D vortex glass
theory or plastic creep theory, we conclude that the vortex motion
above the second peak is plastic when $j\rightarrow 0$ and there
is no vortex glass phase at finite temperatures in Bi-2212. The
peak of S is then explained as the crossover between different
meta-stable vortex states.
\end{abstract}

\end{frontmatter}

\section{Introduction}
One of the interesting phenomenon in the vortex system of high
temperature superconductors (HTS) is the second-peak (SP) effect,
which represents an anomalous increase of critical current with
the increase of magnetic field. For the strong layered
superconductor Bi-2212, it has already been pointed out that the
SP may originate from the phase transition between a quasi-ordered
Bragg glass (BG) and a disordered vortex phase\cite{Giamarchi}.
The existence of BG has been supported by neutron scatting
experiments\cite{neutron}, but it is a long standing question
about what is the nature of the disordered vortex phase in high
field region. Some literatures\cite{VGinBi2212} reported that a 3D
vortex glass (VG) exists above the second peak field and the flux
dynamics is elastic with a diverging barrier in the small current
limit. The vortex glass is assumed to melt into vortex liquid at a
temperature $T_g$\cite{Fisher,Fisher2,Blatter}, around which a
scaling behavior of E-j curves is given. On the other hand it was
predicted from numerical simulation that a plastic deformation
will occur via the proliferation of vortex defects in high field
region\cite{Ryu}. Similarly vortex dislocations are supposed to
exist above the SP field\cite{Giamarchi,Kierfeld}, and Kierfeld
{\it et al.} concluded that the disordered vortex phase and vortex
liquid will be thermodynamically indistinguishable after the
critical endpoint of the first-order melting (FOT)
line\cite{Kierfeld2}. Experimentally Miu {\it et al.}\cite{Miu}
claimed the existence of plastic vortex creep above the second
magnetization peak based on the data measured at a single
temperature T = 25 K. It remains, however, unclear whether this
plastic phase appears in the high field region for all
temperatures. In this paper we will show clear evidence that the
flux motion above the SP field is dominated by plastic motion with
a finite barrier in the small current limit.

\section{Experimental}
Several Bi-2212 single crystals with typical size of
$2\times2mm^2$ and thickness of $20\mu m$ were used in the
experiment. All of them are optimal doped or slightly overdoped
with $T_c= 89 - 91K$ and $\Delta T_c\leq 1K$ determined in
resistive measurement. The samples were measured by a vibrating
sample magnetometer ( VSM 8T, Oxford 3001 ) and a superconducting
quantum interfere device ( SQUID, Quantum Design, MPMS5.5 ). In
all measurements the magnetic field H was applied parallel to the
c-axis and the magnetization data were collected in the field
descending process in order to eliminate the surface barrier. Each
sample shows similar results, therefore we will not distinguish
them for convenience.

\section{Results and Discussion}
\begin{figure}[b]
\includegraphics[scale=1]{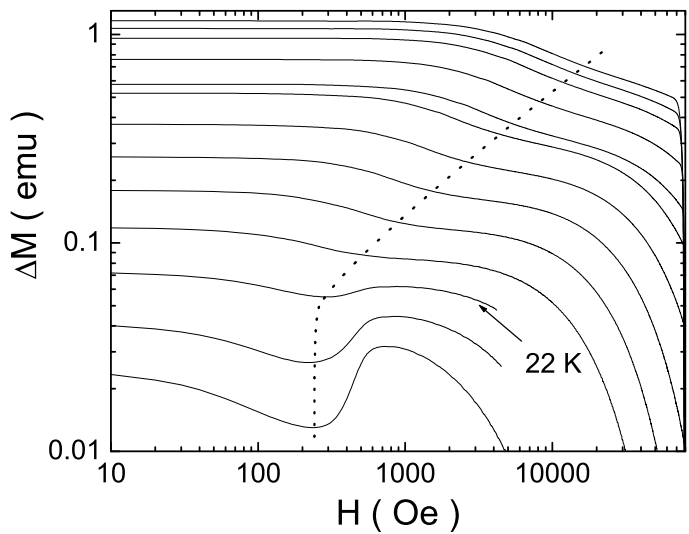}
\caption{ Field dependence of $\Delta M$ at T = 4, 5, 6, 8, 10,
12, 14,16, 18, 20, 22, 24, 26 K from top to bottom. The dotted
line indicates the dip of each curve. The SP is very clear as
pointed by the arrow at 22K.} \label{fig1}
\end{figure}

Fig. \ref{fig1} shows the relationship between the width of the
magnetization-hysteresis-loops (MHL) $\Delta M$ and $H$ in
double-logarithmic coordinates. An interesting feature is that
each curve has a dip, as indicated by the dotted line in Fig. 1.

The dip field decreases with the increase of temperature and
finally drops to the valley field associated with the SP at 22 K
and keeps stable at higher temperatures. Recently we have pointed
out that the SP field reappears and is independent of temperature
well below 20 K\cite{slli}, thus the dips at low temperatures must
have no direct relation with the SP in respect that the dip fields
($> 10^4$ Oe when $T < 10$ K) are much higher than the SP field
(about 280 Oe for this sample).

\begin{figure}[b]
\includegraphics[scale=1]{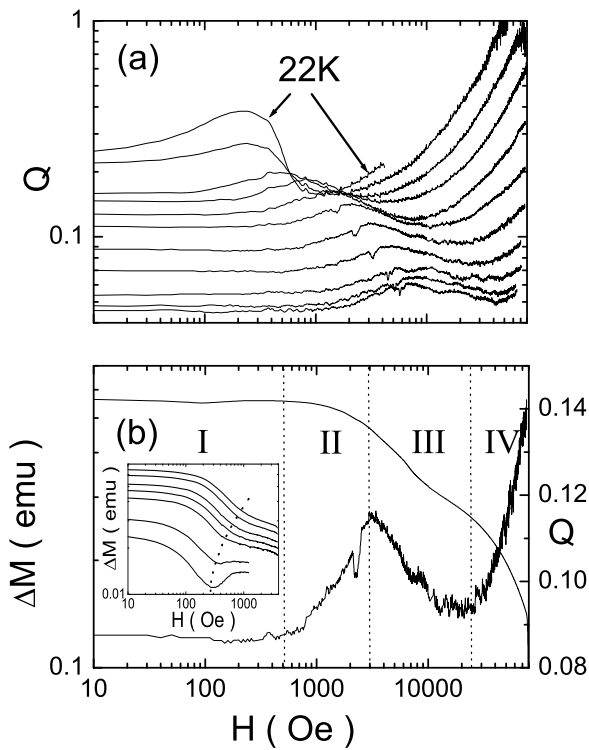}
\caption{ (a)Field dependence of dynamical relaxation
rate Q. The temperatures are the same as in Fig. 1. It is clear
that each curve has a peak and a valley. (b) Field dependence of
$\Delta M$ and Q at T = 10 K. Four regions corresponding to
different vortex dynamical properties are marked by I, II, III and
IV. The inset in (b) gives the $\Delta M - H$ curves at 17 K
measured with different field sweep rates: 100, 50, 20, 10, 5, 1,
0.3 Oe/s. It is clear that the dip has a dynamical feature and
merges to the valley position of SP at higher temperatures. }.
\label{fig2}
\end{figure}

In order to find out the underlying physics corresponding to the
dip on $\Delta M(H)$, we have measured the dynamical relaxation
rate defined as $Q = d(lnj_c)/d(ln(dH/dt))$ by using different
field sweep rates\cite{DynamicalRelaxation}. It has been proved
that $Q$ determined in the dynamical relaxation process is
equivalent to the conventional relaxation rate defined as
$S=-dlnM/dlnt$. The results are given in Fig. \ref{fig2}(a). Just
like the dip on each $\Delta M(H)$ curve, each $Q(H)$ curve shows
a peak. For having a deeper insight to these features, we plot $Q$
and $\Delta M$ vs. $H$ at 10 K together in Fig. \ref{fig2}(b). It
is found that each $Q-H$ curve can be divided into four parts
marked as I, II, III and IV. In region I, both $\Delta M$ and $Q$
are flat, i.e., independent of the magnetic field. In region II,
$Q$ increases monotonously with the increase of magnetic field,
and $\Delta M$ begins to drop. While in region III, $Q$ decreases
with the increase of magnetic field, and the dropping rate of
$\Delta M$ becomes slower and thus results in the dip in MHLs.
Therefore, the dip on MHL is directly related to the peak on the
$Q-H$ curve. The inset of Fig. \ref{fig2}(b) gives the $\Delta M
(H)$ curves of another sample at 17 K and different field sweep
rates. It is clear that the dip field moves to lower value when
the field sweep rate decreases (corresponding to smaller current)
and finally merges to the valley field of SP. Hence both of the
dip on $\Delta M(H)$ and the peak on $Q(H)$ have dynamical
features and should disappear above the SP field in the small
current limit in low temperature region. After decreasing in
region III, $Q$ upturns and increases monotonously with the field
up to the irreversibility field $H_{irr}$ ($Q \approx
1$\cite{Wen97})in region IV. Due to the dynamical feature of the
dip and peak mentioned above, it can be expected that the region
IV will dominate the whole $H-T$ phase diagram above the SP field
in long time or small current limit. After checking about 10
pieces of Bi-2212 single crystals, it is assured that the dip and
peak feature appears in all samples without one exception. In
fact, Cohen {\it et al.} have observed the same dip and peak
phenomena in Bi-2212. They related the peak in $Q$ to the
penetration field\cite{Cohen}. Since the peak can still be found
if we consider only the descending branch of MHLs, which strongly
indicates that there is no relation between the peak and the
penetration field.

\begin{figure}[b]
\includegraphics[scale=1]{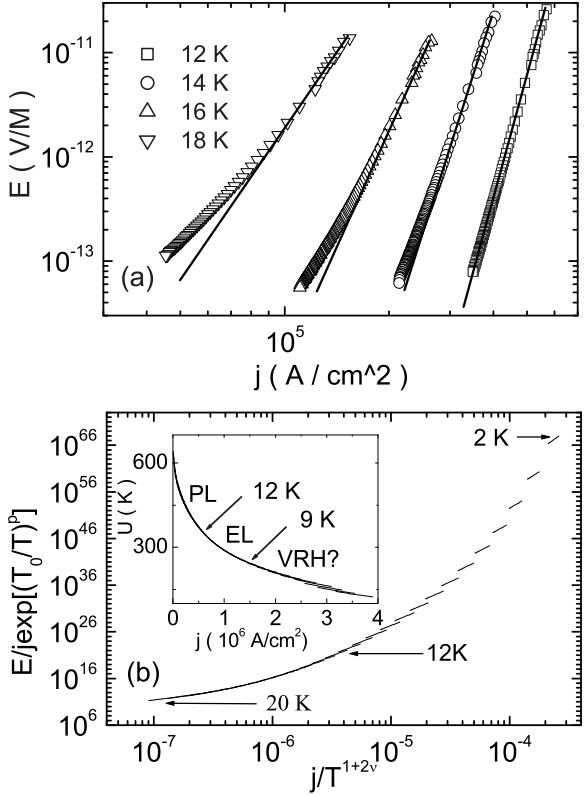}
\caption{(a) $logE-logj$ curves determined from the relaxation
data at H = 1 T and T = 18, 16, 14, 12 K from left to right. The
$T_{irr}$ at 1 T is about 30 K. The linear lines are drawn to help
identifying the curvature of these curves. (b) Scaling the $E-j$
data at H = 5 kOe and T from 5 to 20 K step 0.5 K according to 2D
VG or plastic theory with $T_0 = 350 K$, $\nu_{2D} = 2 $, $ p =
1.2$. The scaling fails below 12 K, as indicated by the arrow. The
inset of (b) gives the U(j) dependence determined by the Maley's
method.} \label{fig3}
\end{figure}

For Bi-2212 single crystals, it was found that the geometrical
barrier plays an important role on the magnetization in the field
ascending process\cite{Geometrical}. The total irreversible
magnetic moment of the sample is contributed by both the surface
shielding current from the geometrical barrier and the bulk
screening current in the field ascending process when the external
field is not high. In this case local magnetic measurement using
either the tiny Hall probe\cite{Zeldov} or magneto-optical
technique\cite{MO} were suggested. While when the external field
is much higher than the second peak field, the contribution from
the bulk current is much larger than that from the geometrical
barrier and a quasi-Bean critical state model is assumed to be
valid\cite{MO}. This is essentially true in the field descending
process when H is decreased from a high field to a field which is
still higher than the second peak field. No superposition of the
magnetic moment due to the bulk current and the surface shielding
current are expected here. As mentioned before our magnetization
relaxation measurement was conducted in the field descending
process with the external field above the second peak field.
Therefore it is justified for us to derive the bulk screening
current density $j$ from the magnetization. It has already been
pointed out that the $E-j$ curve can be obtained by assuming
$M\propto j$ and $dM/dt\propto E$\cite{Ries}. Fig. \ref{fig3}(a)
shows the $E-j$ curves at 1 T based on this method. All the curves
above 12 K are obvious concave, suggesting a finite linear
resistance. Worthy of noting is that the irreversibility
temperature $T_{irr}$ at 1 T is about 30 K. This is a strong
evidence that the vortex motion in region IV is not elastic. In
order to check whether the vortex glass exists below 12 K, a naive
idea is to treat 12 K as the vortex glass melting temperature
$T_g$ and scale the $E-j$ curves using 3D or quasi-2D glass
theory\cite{Yamasaki}. The scaled curve looks not good and the
critical exponents are unreasonably large, e.g., $\nu \sim 3.5$,
which is commonly believed to be less than 2\cite{Blatter}. Hence
we fit it to the 2D VG theory\cite{Fisher2,Fisher3}, which
predicts $T_g = 0$. According to 2D VG theory the correlation
length $\xi_{2D}$ diverges at $T = 0$ K, $\xi _{2D} =
a_0(\varepsilon_0d/k_BT)^{\nu_{2D}}$, where $\nu_{2D}$ and $a_0$
are the 2D VG exponent and the spacing of the vortex lattice
respectively; $\varepsilon_0d$ is the core energy of a vortex
segment of length d. The linear resistivity $\rho_{lin}$ due to
thermal activation is given by $\rho_{lin} \propto
exp(-(T_0/T)^p)$, where $T_0$ is a characteristic temperature of
the same order of magnitude as $\varepsilon_0d$, and
$p=1+\alpha\nu_{2D}$. Therefore we can scale the $E-j$ data as
follows\cite{Wen98},

\begin{equation}
\frac{E}{j}exp{(\frac{T_0}{T})^p} = g (\frac{j}{T^{1+\nu_{2D}}})
\label{scale2D}
\end{equation}

where g is an universal scaling function. The scaled result with
$T_0 = 350 K$, $\nu_{2D} = 2 $, $ p = 1.2$ is given in
Fig.\ref{fig3}(b). It is interesting to note that the values found
here are very close to those found for very thin $YBa_2Cu_3O_7$
films and $Tl_2Ba_2CaCu_2O_8$ thin films at a high magnetic
field.\cite{Wen98} Above 12 K, all data collapse onto one branch,
while the deviation begins at about 12 K. This deviation accords
with the behavior of $\mu$ ( shown below ) which turns from
negative to positive around 12 K. In fact, we can find the same
result if scaling the data based on plastic creep
theory\cite{Reichhardt}. This is not strange if we recall that the
plastic creep theory assumes also a correlation length $\xi$
diverging at 0 K, $\xi \propto T^{-\nu}$, and a linear resistivity
when $j$ approaches zero, $\rho|_{j\rightarrow 0} =
exp(-U_{pl}/k_BT)$, where $U_{pl} \propto (T_c-T)/H^{1/2}$ is the
plastic pinning barrier\cite{Vinokur}. Therefore, we conclude that
the flux motion in region IV is plastic. This indicates that the
vortex glass phase which is characterized by a diverging
activation barrier in the small current limit may not exist in
region IV. Actually, Miu {\it et al.} have also found the plastic
vortex creep above SP, but their claim was made for only a single
temperature ( T = 25 K ).

\begin{figure}[b]
\includegraphics[scale=1]{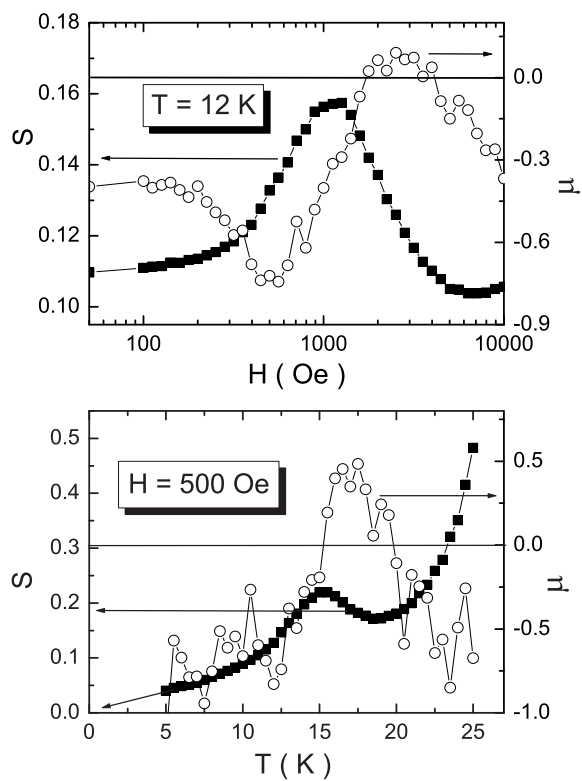}
\caption{The relaxation rate $S$ and the exponent
$\mu$ at (a) $T$ = 12 K and $H$ =10 to 10000 Oe, (b) $H$ = 500 Oe
and $T$ = 5 to 25 K for another sample\cite{slli}. The linear
approach in low temperature region as shown by the arrow in (b)
suggests that there is weak effect of quantum tunnelling creep.}
\label{fig4}
\end{figure}

An important parameter characterizing the vortex dynamics is the
exponent $\mu$, which can be derived from the conventional
magnetic relaxation measurement. Malozemoff {\it et
al.}\cite{Malozemoff} assumed a general pinning barrier
$U=U_0/\mu\cdot[ (j_{c0}/j)^\mu-1]$ from which Thompson {\it et
al.} gave an interpolation formula to obtain $\mu$\cite{Thompson},

\begin{equation}
M(t)^{-\mu}=(M')^{-\mu}+Cln(t), \label{eq1}
\end{equation}

\noindent where $M'$ and $C$ can be treated as constants
approximately. To obtain $\mu$ values we have measured $M$ vs.
time by SQUID and the calculated results at $T = 12$K and various
fields are given in Fig. \ref{fig4}(a), in which the conventional
magnetic relaxation rate $S=-dlnM/dlnt$ is also given. To avoid
the geometrical barrier in the field penetration process, each
measurement was done as following: the field is raised to 5.5 T
and then lowered to the expected field. It is clear that $\mu$ is
negative in region IV. A negative $\mu$ indicates a finite $U$
when $j$ approaches to zero, thus conflicts with the vortex glass
theory.  The $S(T)$ and $\mu (T)$ curves have very similar feature
as $S(H)$ and $\mu (H)$, as shown in Fig. \ref{fig4}(b) for
another sample with $H_p$ around 380 Oe\cite{slli}. The negative
$\mu$ values found here corroborate the picture that the vortex
motion in region IV is plastic.

\begin{figure}[b]
\includegraphics[scale=1]{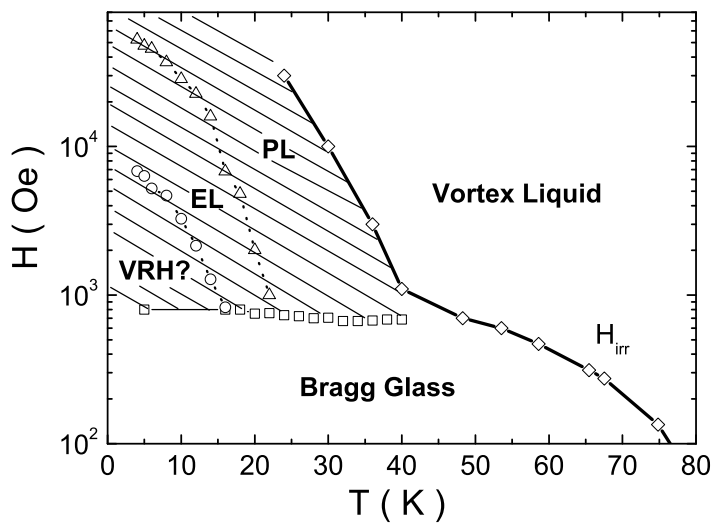}
\caption{ Vortex phase diagram of Bi-2212 single
crystals. The $H_{irr}$ is determined from the deviating point on
the M(T) curves measured in the ZFC-FC processes. The $H_p$ is
taken from our earlier data\cite{slli} and shown as the square
symbol. The boundaries between PL, EL and VRH are determined from
the peak and valley positions of $Q$ in Fig. 2(a). In small
current limit, the EL and VRH region will disappear and the whole
region above SP field becomes plastic. } \label{fig5}
\end{figure}

Next we try to clarify the origin of a positive $\mu$ value and
the drop of $Q$ with increasing field in region III. Joumard {\it
et al.} have measured the ac susceptibility of $(K,Ba)BiO_3$ and
found the elastic (EL) creep in some part of the disordered
phase\cite{Joumard}, which is explained as $R_{\perp} < R_a$,
where $R_{\perp}$ is the dimension of the creeping vortex bundle
and $R_a$ is the mean distance between two dislocations.
Therefore, we attribute the vortex motion in region III to the 2D
EL too. It was predicted that $R_a \sim 1-10 a_0$ above the SP
field\cite{Giamarchi,Kierfeld}, $R_{ \perp} =
a_0(j_{VRH}/j)^{5/8}$ where $j_{VRH}$ is the critical current
separating the 2D EL creep and the variable-range-hopping regions
(VRH). Assuming $R_a\approx 2a_0$, taking $j_{VRH} \approx 1.5
\times 10^5 A/cm^2$, we have $R_a \leq R_{\perp}$ when $j\leq
5\times 10^4  A/cm^2$. It is thus tempting to relate the positive
$\mu$ value as an evidence of 2D EL creep. These regions are
marked on the U(j) curves respecting to different current values
in the insert of Fig.3(b). By further lowering T and/or H
(increasing j) the optimal hopping distance $u$ will be smaller
than $u_{VRH}$ eventually, and the VRH process will take place,
leading to the fast increase of $S$ with H and T in region II. It
is interesting to mention that in $YBa_2Cu_3O_7$ crystals with
columnar defects Thompson {\it et al.}\cite{Thompson2} found the
similar peak on S(T) curve and dip on $j(T)$ curve, which was also
interpreted as the crossover from elastic creep to VRH. It is yet,
however, to be understood why the $\mu$ values determined in
region I and II are negative. There are two possibilities for the
negative $\mu$ values in regions I and II: (1)The total magnetic
moment in the time window we used varies too little for the
program to fit and get a valid and physical meaningful $\mu$ value
( for region II ); (2)A superposition of the bulk current and the
geometrical surface current may occur in region I. Although all
these hypotheses about the dynamical processes in regions I, II
and III need further clarification, it seems quite certain that
the region IV has a feature of plastic creep (PL) and it will
become dominant above the SP field throughout whole temperature
region in the small current limit ( shown by the inset of Fig.2(b)
). Fig. \ref{fig5} gives the phase diagram of Bi-2212. EL and VRH
region will gradually reduce when $j$ decreases, and the whole
region ( shown by the shaded area ) above SP field is dominated by
plastic creep ultimately with $j \rightarrow 0$.

\section{Concluding Remarks}
In conclusion, we have found a general dip on $j-H$ and $j-T$
curves and peak on $S-H$ and $S-T$ curves. They have dynamical
features and are explained as the crossovers between different
vortex phases. Plastic flux motion is found to be dominant far
below the irreversibility line and takes over the whole region
above the SP field in the small current limit.

\label{}
\section{Acknowledgments}

This work is supported by the National Science Foundation of China
(NSFC 19825111, 10274097), the Ministry of Science and Technology
of China ( project: NKBRSF-G1999064602 ), the Knowledge Innovation
Project of Chinese Academy of Sciences. We are grateful for
fruitful discussions with Dr. Sean Ling in Brown University, Rhode
Island, and Dr. Thiery Giamarchi in University of Geneva.



\end{document}